\documentclass[useAMS,usenatbib]{mn2e}

\newcommand{\mr}[1]{\mathrm{#1}}

\usepackage{epsfig}
\renewcommand{\theenumi}{(\roman{enumi})}

\title[Anisotropic mass ejection in binary mergers]
      {Anisotropic mass ejection in binary mergers}
\author[T. Morris and Ph. Podsiadlowski]{T. Morris$^{1,2}$\thanks{E-mail:
tsm@mpa-garching.mpg.de} and Ph. Podsiadlowski$^1$  \\
$^1$Dept.\ of Astrophysics, University of Oxford, OX1 3RH U.K. \\
$^2$Max-Planck Institut f\"ur Astrophysik, Karl-Schwarzschild-Str. 1,
85741 Garching, Germany}
\begin{document}

\date{Accepted 2005 September 23}

\pagerange{\pageref{firstpage}--\pageref{lastpage}} \pubyear{2005}

\maketitle

\label{firstpage}

\begin{abstract}
We investigate the mass loss from a rotationally distorted envelope
following the early, rapid in-spiral of a companion star inside a
common envelope.  For initially wide, massive binaries
($M_1+M_2=20\,M_{\odot},$ $P\sim 10$\,yr), the primary has a convective
envelope at the onset of mass transfer and is able to store much of
the available orbital angular momentum in its expanded
envelope. Three-dimensional SPH calculations show that mass loss is
enhanced at mid-latitudes due to shock reflection from a
torus-shaped outer envelope. Mass ejection in the equatorial
plane is completely suppressed if the shock wave is too weak to
penetrate the outer envelope in the equatorial direction (typically
when the energy deposited in the star is less than about 
one-third of the binding
energy of the envelope). We present a parameter study to show how the
geometry of the ejecta depends on the angular momentum and the energy
deposited in the envelope during a merging event.  Applications to the
nearly axisymmetric, but very non-spherical nebulae around SN\,1987A and
Sheridan 25 are discussed, as well as possible links to RY Scuti and
the Small Magellanic Cloud object R4.
\end{abstract}

\begin{keywords}
hydrodynamics -- binaries: close -- circumstellar matter --
supernovae: individual: SN1987A
\end{keywords}

\section{Introduction}
\begin{figure*}
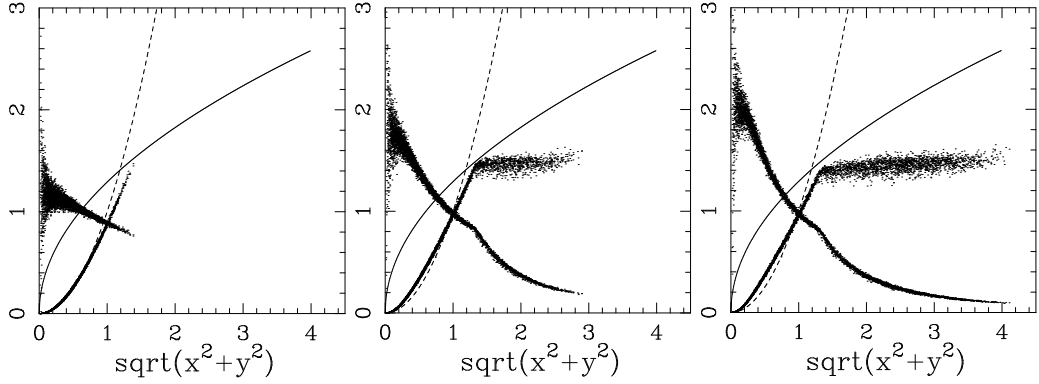

 \centering
  \psfig{file=fig1a.ps, height=4.5cm, angle=270}
  \psfig{file=fig1b.ps, height=4.5cm, angle=270}
  \psfig{file=fig1c.ps, height=4.5cm, angle=270}
  \caption{Particle plots showing specific angular momentum
  (increasing outwards) and angular velocity for the three cases of angular
  momentum $L=0.235,\, 0.588,\, 0.817 \sqrt{GM^3R}$ from left-hand
  panel to right-hand panel.
  The solid curves give the critical specific angular momentum
  $j_c=\sqrt{G(5M/3)R}$ while the dashed curves show the solid body profiles
  $j=\omega r^2$ where $\omega=1$. All quantities are given in code units.}
  \label{fig:ang-part}
\end{figure*}
\begin{figure*}
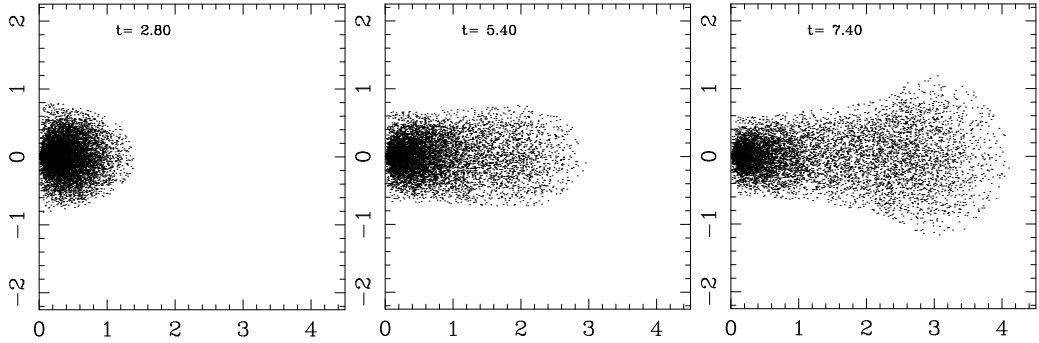

 \centering
  \psfig{file=fig2a.ps, height=4.5cm, angle=270}
  \psfig{file=fig2b.ps, height=4.5cm, angle=270}
  \psfig{file=fig2c.ps, height=4.5cm, angle=270}
  \caption{Distribution of SPH particles in the meridional plane
  immediately after the spin-up for the three cases of angular
  momentum $L=0.235,\, 0.588,\, 0.817 \sqrt{GM^3R}$ from left-hand panel to
  right-hand panel. For other properties see Table \ref{table:env}.}
  \label{fig:env-part}
\end{figure*}
The common-envelope (CE) phase is one of the most important and least
understood phases of stellar evolution. Originally proposed by
\citet{P1976} to explain the origin of short-period binaries with
compact objects, it can also significantly alter the evolution of
systems in which the envelope remains bound, leaving an atypical
single star \citep[see e.g.][]{IP2003}. For example, it is now widely
believed that the unusual properties of the progenitor of SN\,1987A are
due to a binary merger some 20,000\,yr before the explosion
\citep{PI2003}.

In this paper we are interested in the case where the primary
initiates mass transfer either when crossing the Hertzsprung gap
(so-called early case B mass transfer) or later as a red supergiant
(late case B/C mass transfer). In the latter case, the primary has
already developed a deep convective envelope and mass transfer is
dynamically unstable if the mass ratio exceeds a critical value,
leading to a common-envelope and spiral-in phase.

Early case B mass transfer initially occurs on the thermal timescale
of the mass donor and is dynamically stable; but the secondary may not
be able to accrete all of the transferred mass, and this may also lead
to a common-envelope system and possibly the merging of the system
\citep[e.g.][]{Pols1994,WLB2001}.

While the common envelope maintains co-rotation with the embedded binary,
orbital angular momentum is efficiently transferred from the binary orbit
to the envelope, where most of the initial orbital angular momentum
is stored
\begin{equation}
  L_{\mr{orb}}=
  6.60 \times 10^{54} \mathrm{g\,cm}^2 \mathrm{s}^{-1} \,
       A_{2500}^{1/2} \,
       M_{15} \, M_5 \, M_{20}^{-1/2},
  \label{eq:angmom}
\end{equation}
where $A_{2500}$ is the orbital separation in units of
2500\,$R_{\odot}$, $M_{15}$ and $M_{5}$ are the masses of the primary
and the secondary in units of $15\,M_{\odot}$ and $5\,M_{\odot}$ (as
indicated by the subscripts), respectively, and $M_{20}=M_1+M_2$ is
the total mass in units of $20\,M_{\odot}$.  This phase may last
perhaps for a few decades and ends when the envelope becomes
differentially rotating.  The subsequent rapid plunge-in of the
secondary then drives significant envelope expansion and the ejection
of at least some of the envelope
\citep{MM1979,Sa1998,TS2000,Podsi2001,IP2003}.  If the envelope is not
completely ejected in this phase (the case of interest in this study),
the spiral-in continues and now becomes self-regulated where all the
energy dissipated by the further orbital decay is transported to the
surface and radiated away \citep{MM1979,Podsi2001}.  A second phase of
mass loss may result from a nuclear flash that may occur during the
final core merger \citep{IP2003}.

Previous one-dimensional numerical simulations by \citet{Podsi2001}
have shown that significant mass loss may occur even when most of the
envelope remains bound. In the three dimensional models of
\citet{LS1988} and \citet{Sa1998}, most mass loss occurs in the
orbital plane of the binary.  However, these authors considered the case
where most/all of the envelope was ejected. In this study, we consider
the less energetic case appropriate for a merger. As we will show, in
this case mass loss may preferentially occur at mid-latitudes and be
suppressed in the equatorial direction if the energy deposited is less
than about one-third of the binding energy of the envelope.
In section~2 we outline our numerical method  and in section~3 we
present the main results of our study and their dependence on the
input parameters. In section~4 we apply these results
to observed systems, in particular SN\,1987A and Sheridan 25.

\section{Numerical Method}
We model the common envelope as a condensed polytrope with adiabatic
index $\gamma=5/3$ with a central point mass which contains two-fifths of
the system mass\footnote{These parameters were chosen to roughly
represent the inferred properties during the late spiral-in phase for
merger models of SN~1987A, where the core fraction represents the
immersed binary core, consisting of the core of the primary and the
spiraling-in companion.}.  Assuming spherical symmetry and
hydrostatic equilibrium initially, we obtain the radial density
profile by integrating the dimensionless equations,
\begin{equation}
  \frac{\mr{d}\bar{\rho}}{\mr{d}\bar{z}} = -A\frac{\bar{m}}{z^2}
    \bar{\rho}^{2-\gamma}=
    \left(\frac{G}{\gamma
    K}\frac{M_{\mr{core}}^{2-\gamma}}{R_{\mr{core}}^{4-3\gamma}}
    \right) \frac{\bar{m}}{z^2} \bar{\rho}^{2-\gamma},
\end{equation}
\begin{equation}
  \frac{\mr{d}\bar{m}}{\mr{d}z} = 4\pi z^2 \bar{\rho}
\end{equation}
with the inner boundary conditions
\begin{equation}
  \bar{m}(z=1) = 1
\end{equation}
and
\begin{equation}
  \bar{\rho}(z=1) = (\gamma A)^{1/\gamma}\left(
    \frac{3(\delta-1)}{8\pi} \right)^{1/\gamma},
\end{equation}
where $z=r/R_{\mr{core}}$,\, $\bar{m}=M(r)/M_{\mr{core}}$ and
$\rho=\bar{\rho} M_{\star}/R_{\star}^3$. The free parameters
$A,\delta$ are determined from the surface boundary conditions
$\bar{\rho}(z=Z)= 0,\, \bar{m}(z=Z)= M_{\star}/M_{\mr{core}}$ where
$Z=R_{\star}/R_{\mr{core}}$.
For all hydrodynamical simulations presented in this paper, we use the
{\small GADGET} code of \citet{SYW2001}, which implements gravity and gas
dynamics using the smoothed particle hydrodynamics (SPH) method
\citep{Mon1992}. The envelope density is sampled with $10^5$ particles
using a Monte-Carlo method followed by isentropic relaxation to reduce
numerical noise \citep{L1977}. The code units are $M$, the mass of the
envelope and $R$, the initial (non-rotating) stellar radius
(the total stellar mass including the core is $5M/3$). This implies
that time is measured in the code in units of $\sqrt{R^3/GM}$ and
velocity in units of $\sqrt{GM/R}$. Note that this allows simple
re-scaling of the results presented in this paper.

To parametrize the spin-up of the envelope and the energy deposited
by the spiral-in, we define two parameters $\alpha$ and $\beta$, where
$\alpha\equiv \Delta E/ E_{\rm B}$ is the ratio of the energy deposited to the
binding energy of the envelope  and $\beta\equiv L/\sqrt{GM^3R}$ is
a dimensionless measure of of the envelope angular momentum following the
early in-spiral of the secondary.

To spin up the envelope we add angular momentum on a dynamical timescale
using the following recipe: during every fixed timestep
$\Delta t=0.025$, the rotational velocity of each particle is incremented by
an amount
\begin{equation}
  \Delta v_i = r_i \Delta \Omega,
  \hspace{6mm}\mathrm{provided}\hspace{6mm} v_i < \sqrt{-\phi_i},
\end{equation}
i.e. as long as the velocity remains sub-Keplerian.
The angular velocity increment is $\Delta \Omega=0.0093$ 
of the critical velocity at
the surface of the non-rotating envelope ($r=1$ in code units).
If at any time the particle velocity $v_i$ reaches the local Keplerian
velocity, we set $\Delta v_i=0$ thereafter. This leads to solid body
rotation in the inner envelope and a slightly rising specific
angular momentum profile in the outer envelope (see Fig.~\ref{fig:ang-part}).
The spin-up phase is terminated when $\beta$ reaches
0.235, 0.588, 0.817, respectively, for the three cases we consider in this
paper (note that  no particles become super-critical;
see Figure \ref{fig:env-part} and Table \ref{table:env}). These three
values correspond to envelope angular momenta of $2.3, 5.7$ and
$8.0\times 10^{54}\mathrm{\,g\,cm^2\,s^{-1}}$ which is comparable to
the available orbital angular momentum (equation \ref{eq:angmom}).

To simulate the deposition of the energy and the rapid heating of the
envelope during the plunge-in phase, we then add entropy to the inner
envelope [$r<2/15 R$, from calculations discussed in
\citet{IP2003} and \citet{Podsi2001}].  Initially, we consider the case where
the energy is deposited immediately after the envelope has been spun
up. The response of the envelope is then followed for 10\,--\,15 dynamical
timescales after the instantaneous energy deposition, at which point
all the ejected particles are to good approximation on ballistic
trajectories.

Since we do not follow the evolution of the spiraling-in binary
components, we do not encounter resolution problems when the orbital
separation becomes comparable to the SPH smoothing length
\citep{LS1988}; our model is mainly limited by physical approximations
(such as the equation of state; the lack of energy transport) rather
than the numerical resolution (see Appendix A) -- except near the
surface. The steep density profile at the surface is poorly resolved
by SPH particles, and a well known problem of SPH models in this
context (as, e.g., seen in supernova models) is an over-estimate of
the mass contained in low-velocity material.

\section{Results and discussion}
\begin{table}
\begin{center}
    \caption{Properties of the three rotating envelopes immediately
    before the energy deposition for the zero-delay case. All values
    in cgs units are for $M_{\mr{env}}=12\,M_{\odot}$,
    $R=1500\,R_{\odot}$ appropriate to the merger model for
    SN~1987A.}
    \label{table:env}
\begin{tabular}{lccc}
\hline
$T/W$ & 0.039 & 0.117 & 0.151 \\
\hline
$\beta=L/\sqrt{GM^3R}$ & 0.235 & 0.588 & 0.817 \\
Angular momentum $10^{54} \mathrm{\,g\,cm^2\,s^{-1}}$ & 2.3 & 5.7 & 8.0 \\
Mean angular velocity $10^{-8} \mathrm{\,s^{-1}}$ & 3.7 & 2.2 & 1.1 \\
Flattening $R_{\mathrm{eq}}/R_{\mathrm{polar}}$ &1.7 &4.3 & 6.8 \\
Rotation velocity (km/s) & 42 & 21 & 15 \\
Fraction of critical rotation & 0.96 & 0.69 & 0.61\\
Binding energy ($10^{47}$ ergs) & $-6.2$ & $-5.4$ & $-4.5$ \\
\hline
\end{tabular}
\end{center}
\end{table}
\begin{table}
\begin{center}
    \caption{Properties of the rotating envelopes immediately before
    the energy deposition, for $L=0.588\sqrt{GM^3R}$ (top) and
    $L=0.817\sqrt{GM^3R}$ (below) and for three values of the time delay
    following the spin-up of the envelope.}
    \label{table:env-delay}
\begin{tabular}{lccc}
\hline
$t_{\mr{delay}}$ (years) & 0.67 & 1.35 & 2.69 \\
\hline
$T/W$ & 0.090 & 0.078 & 0.071 \\
Angular momentum $10^{54} \mathrm{\,g\,cm^2\, s^{-1}}$ & 5.7 & 5.7 & 5.7 \\
Mean angular velocity $10^{-8} \mathrm{\,s^{-1}}$ & 1.46 & 1.07 & 0.71 \\
Flattening $R_{\mathrm{eq}}/R_{\mathrm{polar}}$ &5.2 &5.4 & 7.0 \\
Rotation velocity (km/s) & 18 & 16 & 13 \\
Fraction of critical rotation & 0.67& 0.63 &0.57 \\
Binding energy ($10^{47}$ ergs) & $-5.2$ & $-5.1$ & $-5.0$ \\
\vspace{2mm}
$T/W$ & 0.121 & 0.104 & 0.088 \\
Angular momentum $10^{54} \mathrm{\,g\,cm^2\, s^{-1}}$ & 8.0 & 8.0 & 8.0 \\
Mean angular velocity $10^{-8} \mathrm{\,s^{-1}}$ & 0.79 & 0.61 & 0.42 \\
Flattening $R_{\mathrm{eq}}/R_{\mathrm{polar}}$ &7.5 &8.3 & 9.6 \\
Rotation velocity (km/s) & 14 & 13 & 11 \\
Fraction of critical rotation & 0.58& 0.56& 0.52 \\
Binding energy ($10^{47}$ ergs) & $-4.3$ & $-4.2$ & $-4.1$ \\
\hline
\end{tabular}
\end{center}
\end{table}
As shown in Figure~1, the addition of substantial angular momentum
strongly distorts the envelope (see, in particular, the
$8\times 10^{54}\mathrm{\,g\,cm^2\,s^{-1}}$ calculation and Table
\ref{table:env}). These profiles
are similar to the $(n,n')=(1.5,0)$ sequence of inviscid polytropic models
calculated in \citet{BO1973}, although our envelopes are more extended
for a given value of $T/W$ (the ratio of kinetic to potential energy)
due to some viscous heating. The outer
envelopes contain $1.7$ and $4\,M_{\odot}$ for $\beta=0.588$ and
$0.817$, respectively, and both have a temperature of $\sim 10^4$\,K.

\begin{figure*}
 \centering
  \epsfig{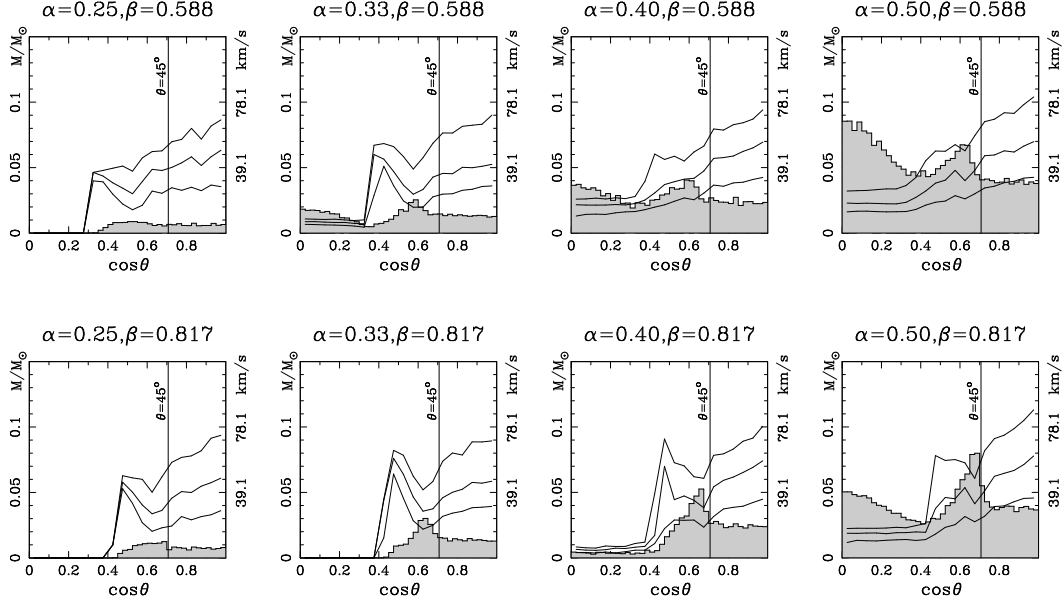}
  \caption{The final amount of mass ejected (left scale) as a function
 of $\cos\theta$ where $\theta$ is the polar angle for different
 values of the energy deposition and different values of the angular
 momentum of the envelope (the values of $\alpha$ and $\beta$ are
 given above each panel; $\alpha=1$ corresponds to an energy of
 $5.4\times 10^{47}$ ergs (for $\beta=0.588$) and $4.5\times 10^{47}$
 ergs (for $\beta=0.817$) while $\beta=1$ corresponds to an angular
 momentum $L_{\mr{env}}= 9.7\times 10^{54} \mathrm{\,g\,cm^2\,
 s^{-1}}$.  The central solid curves show the median velocity of the
 material (right scale) ejected as a function of polar angle, and the
 upper and lower curves give the range of velocities that includes
 50\,\% of the material.}
  \label{fig:nodelay}
\end{figure*}
\begin{figure*}
 \centering
  \psfig{file=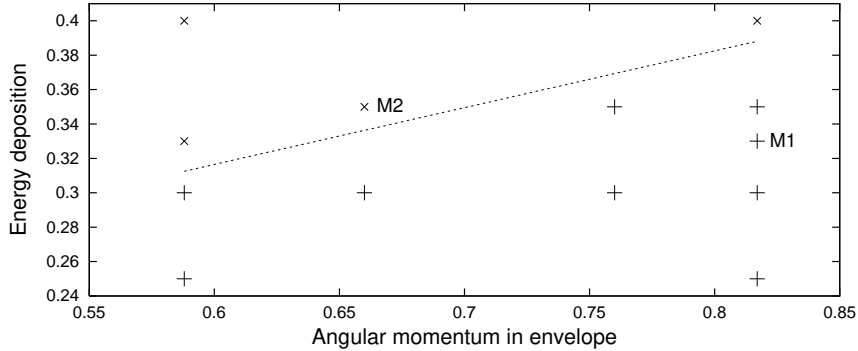, height=11.5cm, angle=270}
  \caption{The ($\beta,\alpha$) plane at $t=10$, for the zero delay case. The
  line separates models with (above) and without (below) equatorial
  mass ejection. Labelled models correspond to models 1 and 2
  discussed in the text.}
  \label{fig:EL-plane}
\end{figure*}

Following the energy deposition, matter is ejected in a very
anisotropic way depending on the rotational distortion and the amount
of energy deposited (i.e. depend on $\alpha$ and $\beta$; see
Fig.~\ref{fig:nodelay}).  Generally envelope material is first ejected
(i.e.\ reaches escape speed) in the polar direction. At low rotation
($\beta=0.235$, not illustrated) mass is also ejected in the
equatorial plane at early times, and the distribution is more or less
spherically symmetric. For larger values of $\beta$, envelope ejection
in the equatorial plane is suppressed for low values of $\alpha$ or
enhanced for large values of $\alpha$. In particular, no mass is lost
in the equatorial plane if the energy fraction $\alpha$ is less than
some critical value $\alpha_c$ which increases with increasing
$\beta$. If no matter is ejected in the equatorial plane, we find a
strong mass excess at mid-latitudes due to shock focusing by the
extended envelope (as discussed further below).

In Sections \ref{sec:model1} and \ref{sec:model2} we discuss two
typical calculations with $\alpha=0.33$, $\beta=0.82$ and $\alpha=0.35$,
$\beta=0.66$,
respectively. In the $\beta=0.82$ case, the critical energy fraction
is $\alpha_c=0.39$ (from Figure \ref{fig:EL-plane}), and therefore no
material is ejected in the equatorial plane. In both these
calculations energy is deposited immediately after the spin-up phase.

In Section \ref{sec:param-study} we discuss the mass loss geometry due
to a variation of $\alpha$ and $\beta$ for different time delays between
the spin-up and energy deposition phase (no time delay and a delay
of 1.35\,yr, respectively; see also Table~2). 
The dynamical time delay is a free
parameter in our model which could be constrained by more detailed models
that follow the spiral-in explicitly.

\subsection{Model 1: $\alpha=0.33, \beta=0.817$}
\label{sec:model1}
\begin{figure}
 \centering
  \psfig{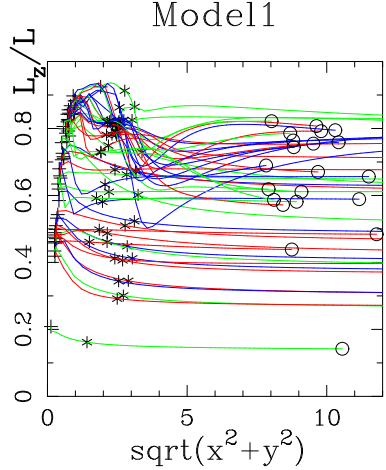}
  \caption{$L_z/L$ as a function of distance from the rotation axis
for selected (ejected) particles in model 1 to illustrate the deviations from
ballistic motion, where $L_z$ is the angular momentum in the z
direction and $L$ the total angular momentum of the particle.  Stars
and circles show particle positions at $t=2$ and $t=15$ (in code
units), respectively. The vertical axis shows $L_z/L$; equatorial
orbits are close to the top of the figure.}
  \label{fig:model1-traj}
\end{figure}
\begin{figure}
 \centering
  \psfig{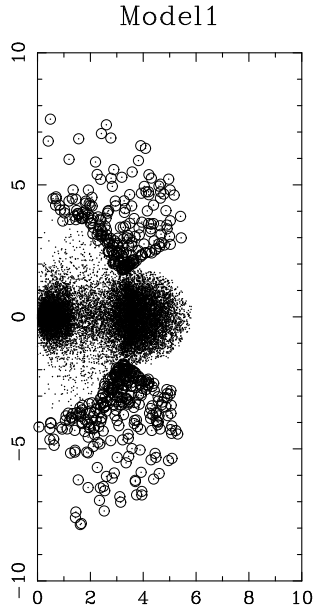}
  \caption{Particle snapshot at $t=3$ showing the deflection of the
    ejected particles (circled) from the outer envelope.}
  \label{fig:partseject}
\end{figure}

In this model, the envelope is significantly distorted by the
rotation (column 3 of Table \ref{table:env}),
and most mass is ejected at a latitude between
$30$ and $40^{\rm o}$. Despite the
reduction in the effective gravity close to the equator, very little
mass is ejected there.
The absence of ejected material in the equatorial plane is relatively
easy to understand from the strong rotational deformation,
since the shock wave cannot penetrate the outer envelope in the
equatorial direction.

However, it is not quite so obvious how the mass enhancement at mid-latitudes
arises. To illustrate the origin of this enhancement, we plot in
Figure \ref{fig:model1-traj} the
evolution of the angular momentum in the z-direction relative
to the total angular momentum for selected particles, which eventually
escape, as a function of distance from the rotation axis
during the early ejection phase. The ratio $L_z/L$ parametrises
the inclination of the orbit of a particle, and its change
shows the effects of the strong pressure gradients that act during
this phase. Particles initially ejected at mid-latitudes move
poleward shortly after ejection, reach a minimum in $L_z/L$ before
$t=3$ (e.g. the blue trajectory in Figure
\ref{fig:model1-traj} which reaches a minimum at $L_z/L=0.5$) and then
evolve towards a ballistic orbit with a lower inclination.

During the strong poleward deflection at $t=2$, a bow shock
forms ahead of the massive outer envelope (see Figure \ref{fig:partseject}),
which deflects particles away from the
equatorial plane. Hence the mass loss poleward of $\theta=30^{\rm o}$
changes from isotropic at $t=2$ to strongly peaked at $t=3$, where
$\theta$ is the co-latitude.

The second deviation in Figure \ref{fig:model1-traj} can be understood
by the reflection of particles from slow-moving material at higher
latitudes, which was ejected slightly earlier. This leads to a
compression of the outflow at mid-latitudes ($L_z/L\sim 0.65$),
increasing the density enhancement. After $t=5$ the
particles follow nearly radial trajectories, although the trajectories
are not yet quite ballistic.
By $t=10$ the flow is almost completely ballistic (as indicated by
the fact that $L_z/L$ remains constant thereafter).

\subsection{Model 2: $\alpha=0.35, \beta=0.66$}
\label{sec:model2}
\begin{figure}
 \centering
  \psfig{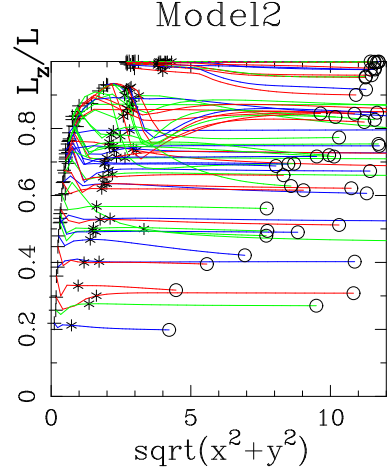}
  \caption{Similar to Figure 2 for model 2.  Stars and circles show
particle positions at $t=2$ and $t=15$ (in code units), respectively.
The vertical axis is $L_z/L$.}
  \label{fig:model2-traj}
\end{figure}
The principal effect of the reduced angular momentum of the envelope
is that the shock wave eventually reaches the surface in the
equatorial region, and that some material is ejected there at low
velocities (similar to the $\alpha =0.33$ and $\beta = 0.588$ model in
Fig.~3 but with a larger excess at mid-latitudes).
The total ejected mass at $t=10$ is
$0.51\,M_{\odot}$, of which $0.04\,M_{\odot}$ is ejected per unit
solid angle in the equatorial plane, a value that is close to the
value one would expect if the ejected mass had spherical symmetry.
Initially the flow is compressed due to the Bjorkman-Cassinelli effect
\citep{BC1993}, but a strong density enhancement of one to two orders of
magnitude does not form due to the fact that our envelope is extended
and mainly supported by thermal pressure.

The mid-latitude enhancement contains a similar amount of mass as
Model 1 discussed above, although its latitude is closer to the
equator. As can be seen in Figure \ref{fig:model2-traj},
the flow is qualitatively similar to Model 1 (Figure
\ref{fig:model1-traj}), but the shock interactions occur at lower
latitudes since the star is less rotationally flattened.  The velocity
profile is similar in the polar region, since almost the same amount
of energy was deposited in the envelope.

As in Model 1, the initial polar enhancement is caused by
the absence of ejected matter elsewhere at early times. This
leads to a large asphericity factor as a result of the small solid
angle into which mass is ejected. Once mass has been ejected
at lower latitudes, the `enhancement' disappears.

\subsection{Parameter study}
\label{sec:param-study}
\begin{figure*}
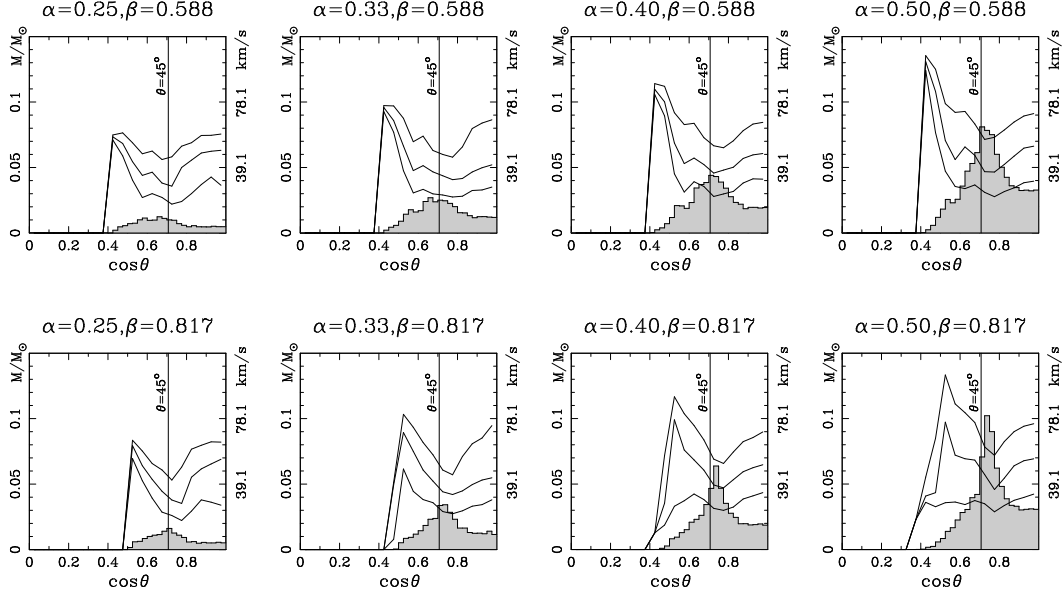

 \centering
  \epsfig{file=fig8a.epsi, height=14cm, angle=90}\vspace{6mm}
  \epsfig{file=fig8b.epsi, height=14cm, angle=90}
  \caption{Properties of the ejected matter, similar to
Figure~\ref{fig:nodelay}, for the case
where the energy is deposited 1.35 years (1.6 dynamical timescales)
after the spin-up of the envelope.}
  \label{fig:delay1}
\end{figure*}

\begin{figure*}
 \centering
  \caption{Total ejected mass for the calculations listed in 
  Table~\ref{table:mass-ejected} after $\simeq 8.2$ years. 
  Darker shading corresponds to longer delay times.}
  \epsfig{file=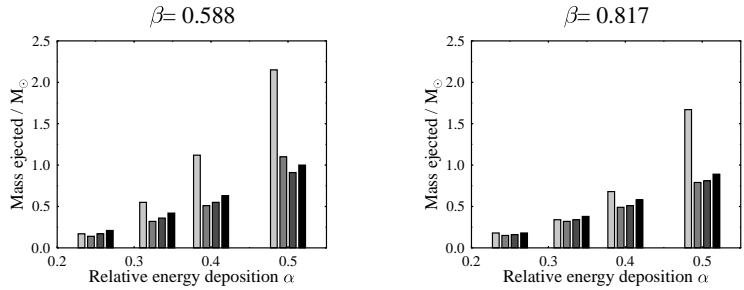, height=4cm, angle=0}
\end{figure*}
\begin{table*}
 \centering
  \caption{Total mass ejected at $t=10\,t_{\mr{dyn}}\simeq
8.2$ years after the entropy
  deposition in the inner envelope, in solar masses. $T$ is the time
  delay in years following the spin-up phase.}
  \label{table:mass-ejected}
  \begin{tabular}{l|c|cccc|cccc}
  \hline
  $\alpha=\frac{E}{E_{\mr{BE}}}$ &
  $\beta=\frac{L}{\sqrt{GM^3R}}= 0.235$ &
  \multicolumn{4}{c}{$\beta= 0.588$} &
  \multicolumn{4}{c}{$\beta= 0.817$} \\
    &$T=0$ &$T=0$& $T=0.67$ & $T=1.35$ & $T=2.69$ & $T=0$& $T=0.67$
	 & $T=1.35$ & $T=2.69$ \\
  \hline
  0.25 & 0.24& 0.17& 0.14& 0.17& 0.21& 0.18& 0.15& 0.16& 0.18 \\
  0.33 & 0.44& 0.55& 0.32& 0.36& 0.42& 0.34& 0.32& 0.34& 0.38 \\
  0.4  & 0.83& 1.12& 0.51& 0.55& 0.63& 0.68& 0.49& 0.51& 0.58 \\
  0.5  & 1.33& 2.15& 1.10& 0.91& 1.00& 1.67& 0.79& 0.81& 0.89 \\
  \hline
  \end{tabular}
\end{table*}

In general, the mass ejected depends on three principal parameters, the
amount of energy deposited ($\Delta E$), the total angular momentum in
the envelope ($L$) and the time delay between the spin-up of the
envelope and the deposition of the energy ($t_{\mr{delay}}$), which
depends on the timescale of the initial spiral-in phase. Variation of
these parameters leads to changes in i) the total ejected mass, ii)
the presence or absence of ejected material in the equatorial plane,
and iii) the strength of the enhancement at mid-latitudes. The peak
mass flux also moves to slightly higher latitudes with increasing
rotation.

In Figure \ref{fig:nodelay} we show the results of a systematic
parameter study.  Each panel shows the geometry of the ejected mass as
a function of $\cos\theta$, where $\theta$ is the polar angle,
once the ejecta are expanding ballistically
on radial trajectories. The histograms show the mass ejected divided
into bins of constant $\mr{d}\cos\theta$ (each bin subtends
$\pi/10$ sr), and the individual curves show the
velocity distribution at each angle (the central curve gives
the median velocity, and the upper and lower curves give the
velocity range which includes 50\,\% of the ejected matter).
The velocity and angular momentum scale according to
\begin{equation}
  v_{\mathrm{cr}}\sim 39\, \mr{km}\, \mr{s}^{-1}\left( \frac{M}{12M_{\odot}}
    \frac{1500 R_{\odot}}{R} \right)^{1/2},
\end{equation}
\begin{equation}
    L_{\mr{env}}= \beta\, 9.7\times 10^{54}\mathrm{\,g\,cm^2\, s^{-1}}
    \left( \frac{M}{12M_{\odot}} \right)^{3/2}
    \left( \frac{R}{1500 R_{\odot}} \right)^{1/2}.
\end{equation}

The peak at mid-latitudes, which is only present for sufficiently
distorted envelopes with $\beta>0.5$, remains up to
$\alpha= 1$.
The mass excess at mid-latitudes becomes more pronounced as the
angular momentum increases. The velocity of the equatorial material,
when present, is typically a factor of 3\,--\,5 lower than the velocities at
mid-latitudes, though this difference is reduced as the deposited energy
is increased. The critical $\alpha$ below which no mass is lost
in the equatorial plane increases with rotation rate (see Figure
\ref{fig:EL-plane}), since the envelope is more extended.

The trends remain very similar, when we introduce a time delay between
the spin-up phase and the energy deposition phase (see
Figure \ref{fig:delay1}), except that the loss of material in the
equatorial plane is now further impeded by the massive extended envelope,
and the mid-latitude enhancement may be much stronger (see, for example, the
$\alpha=0.5$, $\beta=0.817$, $t_{\mr{delay}}=1.35\,\mr{yr}$ case).
The peak has moved to
slightly higher latitudes compared to the zero delay case, but this is
a small effect. During the time delay the outer envelope expands by
a factor $\sim 1.5$ with a corresponding increase in the critical energy for
equatorial mass ejection. Hence increasing the energy
drives more mass from each surface element but has little effect on the
geometry. The total mass ejected for each value of the energy
deposition $\alpha$ and time delay is listed in Table~3 and
illustrated in Figure~9.

\subsection{Summary}
Mass ejection during a common-envelope phase leading to the complete
merger of a binary system preferentially occurs at mid-latitudes due
to shock reflection from an outer envelope containing a significant
fraction of the initial orbital angular momentum. If the energy
deposited is less than about one-third of the binding energy of the envelope
no material is lost in the equatorial plane, in contrast to earlier
models which considered the higher energy case appropriate for CE
ejection \citep{LS1988,Sa1998}.

Although our results cannot be directly compared to those of \citet{Sa1998}
since we do not model the in-spiral of the secondary, a number of
similarities are apparent. The peak energy dissipation occurs when the
orbital separation is approximately one-tenth of its initial value and
the red
giant envelope is already aspherical which is consistent with our
prescription. Most of the mass loss ($\sim 0.3M_{\mr{env}}$) is
concentrated in the equatorial plane in this more energetic case. The
equivalent energy parameter is $\alpha= \Delta E/ E_{\mr{B}}\approx
1.2$ whereas we consider the range $\alpha=0.25-0.5$.

\section{Applications to SN1987A and Sheridan 25}
\label{sec:applications}

\begin{table*}
 \centering
 \begin{minipage}{150mm}
  \caption{Observed properties of circumstellar material around
  massive stars. Bracketed periods denote systems which are too wide
  to undergo binary interactions. The properties of the material in
  the equatorial torus are given in the last four columns, where $R$,
  $v$ and $t$ give the radius, expansion velocity and dynamical age,
  respectively.} 
  \label{table:nebulae}
  \begin{tabular}{llcccccc}
  \hline
  Star & Spectral type & Period & Luminosity
  & \multicolumn{4}{c}{Equatorial torus} \\
    & & & $L/L_{\odot}$
  & Density (cm$^{-3}$) or mass & $R$ ($10^{17}$ cm) & $v$ (km/s) & t
  (yrs) \\
  \hline
  Sk $-69^{\circ}202$ & B3 Ia & - & $10^{5.2}$ & $n_e\sim 2-5\times 10^{4}$ &
  $6.23\pm 0.08^a$ & 10.3 & 19,000 \\
  Sher 25 & B1.5 Iab & -& $10^{5.9}$& $0.01-0.1M_{\odot}$ & 6.2 & 20 & 6,600 \\
  RY Scuti & O6.5 + O9.5$^b$ & 11.3d & $\sim10^6$
    & $1M_{\odot}$? & 0.2 & 50 & 130 \\
  $\eta$ Car & ? Ia & (5.5yr) & $10^{6.7}$ &
  $3-15M_{\odot}^c$
    & 0.7 & 50& $\sim 1,000$ \\
  R4$^d$ & B[e] + A & (21.3yr) & $10^5+10^{4.2}$
  & ? & ? & ? & $\sim 10,000$ \\
  \hline
  \end{tabular}
  Note: Data taken from
  $^a$\citet{P2004}, $^b$\citet{SGG2001}, $^c$\citet{Mor1999,SGH3M22003} 
  and $^d$\citet{ZKWS1996}.
\end{minipage}
\end{table*}
\begin{figure*}
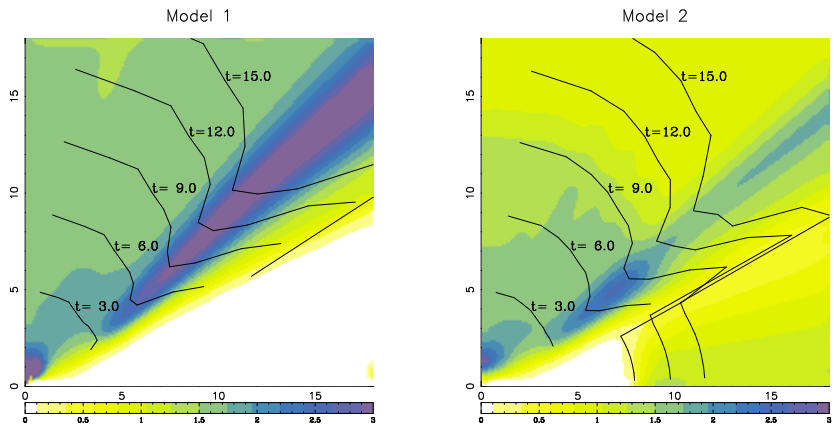

 \centering
  \psfig{file=fig10a.ps, height=5.5cm, angle=0} \hspace{10mm}
  \psfig{file=fig10b.ps, height=5.5cm, angle=0}
  \caption{Mass enhancements in the ejecta flow, corresponding to
  merger models for SN\,1987A (left-hand panel, Model 1 from section~3.1 with
$\alpha = 0.33$ and $\beta = 0.817$) and Sheridan 25 (right-hand panel; Model 2
from section~3.2 with $\alpha = 0.35$ and $\beta = 0.665$).
The solid curves give the locations that contain 50\,\% of the
mass ejected at a particular solid angle at the time as indicated (in
code units).}
  \label{fig:massenh}
\end{figure*}
Many non-spherical nebulae are axially symmetric, which has been interpreted
as evidence for rapid rotation, possibly as a result of binary
interactions. A notable example is the mysterious nebula surrounding the
supernova SN\,1987A, which consists of
three roughly parallel rings, one centered on the
supernova and the other two displaced by $\sim 1$ arc-second to
either side \citep{B1995}. The supernova itself was anomalous
in several other respects \citep[see][and references
  therein]{Podsi1992}, which are most consistent with
a binary merger some 20,000 years before the supernova event
\citep{Podsi1992,PI2003}.
In particular, the blue supergiant progenitor and the chemical anomalies
in the inner ring can easily be explained as the result
of the dredge-up of core material in the final stage of the
merger \citep{IP2003}.

The fast wind ($\dot{M}\sim 10^{-7}M_{\odot}\,\mr{yr}^{-1},
v_{\infty}\sim 500\,\mr{km/s}$) of such a blue supergiant will sweep
up and enhance
any structures already present in the ejecta, leading to an axially
symmetric but highly aspherical nebula.

\subsection{Late case B/C merger: SN 1987A}
During late He shell burning the
primary will be a red supergiant with a critical surface rotation velocity of
\begin{equation}
  v_{\mathrm{cr}}\sim 45\, \mr{km}\, \mr{s}^{-1}\left( \frac{M}{15M_{\odot}}
    \frac{1500 R_{\odot}}{R} \right)^{1/2}
\end{equation}
which is comparable to the velocities observed in the SN\,1987A nebula
(see Table \ref{table:nebulae}). The latitude-dependence of the
ejected material of Model 1, shown in Figure \ref{fig:massenh},
is characterised by a strong enhancement at mid-latitudes, while
no material is lost in the equatorial region. The following features
of the nebula may therefore be understood:
\begin{enumerate}
  \item \textit{Strong mass enhancement in the outer rings}.  The
    outer rings (ORs) are a real density enhancement ($100\times$ the
    ambient value) and are not simply due to limb brightening of an
    hourglass structure \citep{B1995}.  In our model the ORs result
    from wind-driven pressure gradients in the seed structures which
    directly result form the anisotropic ejection of material during
    the merger phase.  Previous models based only on equatorial
    density enhancements in the pre-existing material have been unable
    to explain the high density in the ORs \citep[e.g.][]{MA1995}.
  \item \textit{Displacement of the outer rings relative to the inner
    ring}.  We favour a model in which the inner ring originates a few
    1000 years {\em after} the merger event, in a rotation-enforced
    outflow during contraction on the post-merger blue loop (Heger \&
    Langer 1998; also see Collins et al.\ 1999).  The relative
    displacement of the outer rings can be understood if the mass
    ejection during the merger event itself was slightly asymmetric,
    perhaps due to a non-axisymmetric pulsational instability in the
    envelope, which gives the ejecta a velocity of $\sim
    2\,$km\,s$^{-1}$ relative to the merger remnant. Hence the
    wind-driven pressure gradients are no longer axisymmetric and the
    planes\footnote{The outer rings are still approximately planar in
    this case.} in which the outer rings lie will be slightly inclined
    with respect to the plane of the inner ring. This would explain
    both the offset of the outer rings and their shape, which is
    noticeably non-elliptical in projection.
  \item \textit{North/South asymmetry}.
    Since the planes of the outer rings are no longer parallel to one
    another, the Southern outer ring is observed
    closer to face-on in projection than the Northern one.
\end{enumerate}

\subsection{Case B merger: Sher 25}
Mass loss during common-envelope evolution may explain the broadly
similar structures seen around other luminous stars, listed in Table
\ref{table:nebulae}. Of these, the nebula around the B1.5 supergiant
Sheridan 25 shows the most compelling similarities, since it has an
equatorial ring of at least $0.01-0.10M_{\odot}$ \citep{B1997b} and
polar lobes, each containing $\sim 0.25M_{\odot}$. It is in a post
main sequence, though probably pre-red supergiant, 
evolutionary state \citep[with
N/C $\sim 26$, N/O $=0.36$,][]{Sm2002}.  The dynamical age of the
nebula has been estimated to be around 6,000 years. The observed
velocities suggest an envelope radius of $\sim 300R_{\odot}$ at the
time of ejection, corresponding to a merger during the Hertzsprung gap
crossing.

Although the density structure of the common envelope will
differ from that of a $\gamma=5/3$ condensed polytrope, the results
may still be applicable if the envelope is aspherical. We suggest a
model with equatorial mass loss during the merger, such as Model 2
discussed above (see right-hand panel of Figure \ref{fig:massenh}),
since the envelope cannot store enough angular momentum to generate a
significant post-merger equatorial outflow. Hence an asymmetry during
the merger will displace both the equatorial and polar material from
the site of the merged star. The equatorial ring is offset from Sher
25 by $0.05-0.1$ pc \citep{B1997a}, which is consistent with this model.

\subsection{Conclusions}
The three principal anomalous features of the supernova SN1987A,
viz. its blue supergiant progenitor, its over-abundance of certain
elements, notably He, and the presence of highly structured
circumstellar material, are all consistent with a binary merger some
20,000 years before the explosion. In this paper, we have demonstrated
how density enhancements at mid-latitudes arise during mass ejection
from a rotationally distorted star.
Subsequent interaction with the fast wind of the blue supergiant prior
to the supernova \citep[cf.][]{BL1993}
then leads to the formation of
the outer rings with a density enhancement of a factor of 150, in
calculations of \citet{MP2005} which will be further discussed in a
future paper.

Similarly, the nebula around Sher 25 may be explained by a binary
merger following a CE phase during the crossing of the
Hertzsprung gap by the primary. One notable difference in this case is
that the
equatorial ring likely originates during the merger, which is
consistent with observations showing that the centre of the equatorial
ring is displaced by some 0.05 pc from Sher 25. Future observations of
the rotation rate of Sher 25 would help to confirm this model.

Although the nebula around the more massive system RY Scuti (O9.5 +
O6.5) appears similar, its evolution is probably somewhat
different. Data from the Keck telescope and the HST,
discussed in \citet{SGH1999} and \citet{SGG2001}, show a massive
equatorial dust torus and two narrow rings symmetrical about the
equatorial plane, with a dynamical age of 120 years. The massive torus
probably originated during thermal timescale mass outflow from the
outer Lagrangian L2 point which is still occurring today, albeit at
the much lower rate of $\sim 5\times 10^{-5}M_{\odot}\,
\mr{yr}^{-1}$. Subsequent deflection of the fast wind in a manner
somewhat analogous to the horseshoe model of \citet{Soker1999} may explain
the origin of the two narrow rings.

\citet{PNLSLC2000} have suggested that the B[e] component of the
spectroscopic binary R4 in the LMC could be the result of a Case B/C
merger, i.e.\ the system would originally have been a triple system,
where the companion A star now serves as an astronomical clock and
indicates that the present primary has lost $\ga 40$\,\% of its
zero-age main-sequence (ZAMS)
mass \citep{ZKWS1996}. Since the companion is already a post-main sequence A
supergiant, the pre-merger components cannot be much more massive,
e.g. $12-15\,M_{\odot}$ for the original primary. In addition,
\citet{PNLSLC2000} find evidence of CNO-processed circumstellar
material with a dynamical age of $\sim10^4$ years which is consistent
with ejection during a binary merger.  Direct images of the nebula
would be of great benefit to understanding this system.

Bipolarity is also common in observations of planetary
nebulae. Asymmetric mass loss during a common-envelope phase (with or
without a merger) provides physical motivation for the equatorial
density enhancement functions proposed by \citet{IBP1989} and \citet{LM1991}
\citep[see also][]{F1999}. We speculate that the homunculus nebula around
$\eta$ Carinae may also have originated during a common-envelope phase
since its kinetic energy is $\sim 10^{50}$\,ergs \citep{SGH3M22003},
comparable to the luminous energy of the $1840-1860$ outburst. Both the
mass loss rate in the stellar wind ($1.6\times
10^{-3}M_{\odot}\,\mr{yr}^{-1}$) and its latitude dependence suggest rapid
rotation of the central star \citep{SDG2003,vBKS2003,ALM2004}. In the merger
scenario $\eta$ Carinae was originally a triple system in which the closer
components ($P\sim 30$\,d) merged 150\,yr ago to leave the present
companion in an eccentric 5.5-yr orbit.

\section*{Acknowledgements}
The computations reported here were performed using the UK
Astrophysical Fluids Facility (UKAFF). The authors would also
like to thank N. Langer for stimulating discussions.

\bsp
\label{lastpage}

\begin{appendix}
\section{Resolution Study}
\begin{figure}
 \centering
  \psfig{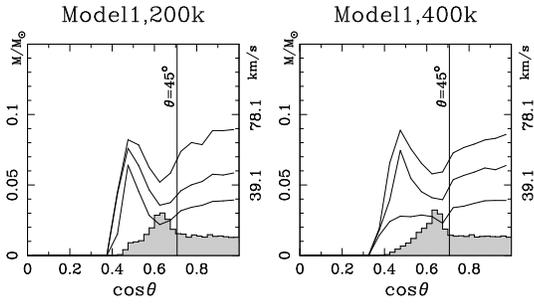}
  \caption{Comparison between two calculations of Model 1, with
  $2\times 10^5$ particles (left-hand panel) and $4\times 10^5$ particles
  (right-hand panel).}
  \label{fig:resstudy}
\end{figure}
\end{appendix}
Figure \ref{fig:resstudy} shows that increasing the number of
particles by a factor of two does not significantly change the
geometry of the ejecta. This implies that our calculations have
converged numerically. The calculation is limited by the physical
approximations such as the assumption of a polytropic equation of state.
\end{document}